\documentclass[11pt]{article}
\usepackage{amsmath,amssymb,color,epsfig,cite}
\usepackage{graphicx}
\usepackage{subfigure}
\usepackage{setspace}

\usepackage{amsfonts}
\newcommand{\hoch}[1]{$\, ^{#1}$}

\makeatletter
\@addtoreset{equation}{section}
\makeatother

\newcommand{\be}{\begin{equation}}
\newcommand{\ee}{\end{equation}}
\newcommand{\bea}{\setlength\arraycolsep{2pt} \begin{eqnarray}}
\newcommand{\eea}{\end{eqnarray}}
\newcommand{\nn}{\nonumber}

\def\ft#1#2{{\textstyle{\frac{\scriptstyle #1}{\scriptstyle #2} } }}
\def\fft#1#2{{\frac{#1}{#2}}}

\def\0{{\sst{(0)}}}
\def\1{{\sst{(1)}}}
\def\2{{\sst{(2)}}}
\def\3{{\sst{(3)}}}
\def\4{{\sst{(4)}}}
\def\5{{\sst{(5)}}}
\def\6{{\sst{(6)}}}
\def\7{{\sst{(7)}}}
\def\8{{\sst{(8)}}}
\def\sst#1{{\scriptscriptstyle #1}}
\def\oneone{\rlap 1\mkern4mu{\rm l}}
\def\ep{{\epsilon}}

\def\wtd{\widetilde}

\def\cA{{{\cal A}}}
\def\cF{{{\cal F}}}

\def\R{{\mathbb R}}

\def\vp{\varphi}
\def\vep{\varepsilon}

\begin{document}

\begin{flushright}
\hfill{MI-TH-1876}

\end{flushright}

\begin{center}
{\large {\bf  On Pauli Reductions of Supergravities in Six and Five Dimensions}}

\vspace{15pt}
{\large\bf  Arash Azizi\hoch{1} and C.N. Pope\hoch{1,2} }

\vspace{15pt}

\hoch{1}{\it George P. \& Cynthia Woods Mitchell  Institute
for Fundamental Physics and Astronomy,\\
Texas A\&M University, College Station, TX 77843, USA}

\vspace{10pt}

\hoch{2}{\it DAMTP, Centre for Mathematical Sciences,
 Cambridge University,\\  Wilberforce Road, Cambridge CB3 OWA, UK}

\vspace{30pt}

\underline{ABSTRACT}

\end{center}

  The dimensional reduction of a generic theory on a curved internal space 
such as a sphere does not admit a consistent truncation to a finite set
of fields that includes the Yang-Mills gauge bosons of the isometry group.
In rare cases, for example the $S^7$ reduction of eleven-dimensional
supergravity, such a consistent ``Pauli reduction'' does exist. In this paper
we study this existence 
question in two examples of $S^2$ reductions of supergravities.
We do this by making use of a relation between certain $S^2$ reductions and
 group manifold $S^3=SU(2)$ reductions of a theory in one dimension higher.
By this means we establish the non-existence of a consistent $S^2$ 
Pauli reduction of five-dimensional minimal supergravity. We also show that
a previously-discovered consistent Pauli reduction of six-dimensional
Salam-Sezgin supergravity can be elegantly understood via a group-manifold
reduction from seven dimensions.

\vfill {\footnotesize Emails: sazizi@physics.tamu.edu\ \ \
pope@physics.tamu.edu}

\thispagestyle{empty}

\pagebreak




\section{Introduction}

   The idea of using a dimensional reduction on a curved space such as a
sphere in order to obtain a lower-dimensional theory with non-abelian
gauge symmetries has a long history, which seems to have originated with
an unpublished communication by Pauli, in 1952 \cite{strau,oraf}.  His
idea was that by dimensionally reducing six-dimensional Einstein gravity
on the 2-sphere, one might obtain a four-dimensional theory describing
gravity coupled to gauge fields that we would now call $SO(3)$ Yang-Mills.
Pauli realised, however, that there were difficulties with this idea, which
we can recognise as being inconsistencies in the higher-dimensional
components of the Einstein equations.  This is in fact a classic example of
an {\it inconsistent truncation} in Kaluza-Klein.   

   One can always,
of course, make a consistent Kaluza-Klein ``reduction'' in which one
simply expands all the higher-dimensional fields in terms of the 
complete set of harmonics (spherical harmonics, in the case of $S^2$), and
retains the entire infinite towers of lower-dimensional massless and massive
fields.  The consistency issue arises if one wants to set all
except a subset of the lower-dimensional fields to zero.  For this
to be a consistent truncation, it is necessary that the full set of
equations of motion for the infinite towers of fields should be compatible
with setting to zero the fields one wishes to truncate.  The danger is that,
because of the non-linear nature of the theory, the fields one is 
retaining might act as sources for the fields one
wishes to set to zero.  This is exactly what happens in the case Pauli
considered; the Yang-Mills fields one wishes to retain actually act as
sources for a set of massive spin-2 fields that one wishes to set to
zero.  From the higher-dimensional point of view, this reflects itself as
an inconsistency between the truncation ansatz and the higher-dimensional
components of the Einstein equations.

    The inconsistency of the truncation to a finite number of lower-dimensional
fields that include gravity and the Yang-Mills fields of the isometry group
of the compactifying manifold is not restricted to the case where
one starts from pure higher-dimensional gravity.  It is in fact what happens
generically when compactifying any higher-dimensional theory that includes
gravity.  There are, however, some notable exceptions, of which the most
celebrated is the reduction of eleven-dimensional supergravity on the
7-sphere.  In this case, the inconsistencies that would arise if 
eleven-dimensional pure gravity alone were reduced on $S^7$ are removed, 
because of further contributions in the higher-dimensional equations
of motion coming from the 3-form potential.  The consistency
of this remarkable reduction was demonstrated in \cite{dewnic}.  
The exceptional situations
where a consistent reduction on a curved internal manifold $M$, retaining a
finite number of fields including gravity and the Yang-Mills gauge fields
of the isometry group of $M$, have been called consistent 
{\it Pauli reductions} \cite{paulired}.  
Other examples include the consistent Pauli reduction of eleven-dimensional
supergravity on $S^4$ \cite{vann1,vann2,cvluposatr}, 
the consistent Pauli reduction of
type IIB supergravity on $S^5$ \cite{baghohsam}, and the 
consistent Pauli reduction of the
$D$-dimensional bosonic string on any group manifold $G$ \cite{Gred}.  In this
latter case, we emphasise that the Pauli reduction on the group manifold
$G$ yields the Yang-Mills gauge bosons of the full isometry group 
$G\times G$. 

   Establishing the consistency of the Pauli reductions in all the
above examples is highly non-trivial.  There is no simple group-theoretic
argument for why the truncation to the finite subset of lower-dimensional
fields should be consistent. This is in contrast to the very well
understood class of ``group-manifold reductions'' on $G$, in which 
the infinite towers of lower-dimensional fields are truncated to just
those that are singlets under the right action $G_R$ 
(or, equivalently instead, the
left action $G_L$) of $G$.  These reductions were first described by DeWitt 
\cite{DeWitt}, and have been named DeWitt reductions in \cite{paulired}. 
The consistency of the truncation follows trivially from the fact that 
the retained fields, which are singlets under $G_R$, clearly cannot act as 
sources for the non-singlet fields that have all been set to zero.  

   In this paper, we shall investigate the consistent Pauli reduction
on the 2-sphere of certain five-dimensional and six-dimensional 
supergravity theories. A common theme in both of these examples will
be our employment of the observation made in \cite{paulired} that in certain
cases one can construct an $S^2$ Pauli reduction of a particular 
$D$-dimensional theory if that theory can itself be obtained from 
a circle reduction of a $(D+1)$-dimensional theory.  If one starts
by constructing the DeWitt $SU(2)$ reduction of the $(D+1)$-dimensional 
theory, and makes a circle reduction on the Hopf fibres of $SU(2)$
viewed as a $U(1)$ bundle over $S^2$, then one obtains a (necessarily
consistent) $S^2$ Pauli reduction of the $D$-dimensional theory.  In the
two examples that we shall be considering in this paper, a further 
complication arises because the $D$-dimensional theories of interest
to us are in fact themselves {\it consistent truncations} of the
circle reductions of the
parent $(D+1)$-dimensional theories.  That is to say, the truncation of
fields can be performed in a $D$-dimensionally covariant way, but not
in a $(D+1)$-dimensionally covariant way.  This imposes a non-trivial
further condition if one wishes to construct the $S^2$ Pauli reduction
in the way we have described, since the additional field truncation in
$D$ dimensions may not be compatible with the Hopf bundle structure.  In fact
in one of our two examples the Hopf construction will fail for this reason, 
while in the other example it succeeds.

   Our first example is motivated by the well-known observation that the 
minimal five-dimensional supergravity, whose bosonic Lagrangian is
\be
{\cal L}_5 = R {*\oneone} - \ft12 {*F_\2}\wedge F_\2 - \ft1{3\sqrt3}\,
  F_\2\wedge F_\2\wedge A_\1\,,\label{d5minlag}
\ee
where $F_\2=dA_\1$,
is very closely parallel to eleven-dimensional supergravity, whose
bosonic Lagrangian is
\be
{\cal L}_{11} = R {*\oneone} - \ft12 {*F_\4}\wedge F_\4 - 
   \ft16\, F_\4\wedge F_\4\wedge A_\3\,,\label{d11lag}
\ee
where $F_\4=dA_\3$.  One is then tempted to conjecture that what works
for $S^4$ or $S^7$ reductions of eleven-dimensional supergravity
would work also for $S^2$ or $S^3$ reductions of the minimal five-dimensional
supergravity.  In particular, one might expect that a consistent $S^2$ 
Pauli reduction of the minimal five-dimensional theory should be possible.

  We shall study this question by making use of the construction developed
in \cite{paulired}, which we described above.  Thus we shall take as
our starting point the minimal supergravity of six dimensions, whose
bosonic sector comprises gravity and a self-dual 3-form.  The circle
reduction of this theory gives minimal five-dimensional supergravity coupled
to a single vector multiplet.  Thus we are able to construct a consistent $S^2$
Pauli reduction of this five-dimensional theory.   Now the minimal 
five-dimensional supergravity itself arises as a consistent truncation of 
this theory, in which the
vector multiplet is set to zero.  However, as we indicated above, 
it is not guaranteed that imposing this 
truncation of the five-dimensional theory is itself compatible with the
previously-established consistent $S^2$ Pauli reduction, and in fact 
we are able to show that the two conditions are inconsistent in this 
example. Although this does not constitute a complete proof that there
exists no possible $S^2$ Pauli reduction of the five-dimensional
minimal supergravity , the fact that this natural way to try to
construct such a reduction fails is strongly suggestive.  In fact other,
direct, attempts to construct a consistent $S^2$ Pauli reduction of the
five-dimensional minimal supergravity have also been unsuccessful.

  Our second example is provided by the six-dimensional gauged 
supergravity of Salam and Sezgin \cite{salsez}.  This has the intriguing
feature that it admits a supersymmetric $S^2\times$(Minkowski)$_4$ vacuum.  
Furthermore, it was shown in \cite{gibpopss} that there is in fact a consistent 
$S^2$ Pauli reduction of the Salam-Sezgin model, which yields 
a four-dimensional supergravity with $SU(2)$ Yang-Mills fields originating
from the isometry group of the 2-sphere, and whose Minkowski vacuum 
corresponds to the six-dimensional $S^2\times$(Minkowski)$_4$ vacuum found
in \cite{salsez}.  Although it is a much simpler example
than the $S^7$ or $S^4$ reductions of $D=11$ supergravity, the underlying
reasons for the consistency of this $S^2$ Pauli reduction are at present
equally mysterious.  

In this paper, we examine whether it is possible to 
reconstruct the $S^2$ Pauli reduction by exploiting the fact that the 
Salam-Sezgin supergravity can itself be obtained from a circle reduction
of a seven-dimensional supergravity.  Such an
embedding of the Salam-Sezgin theory was obtained in \cite{cvgiposs}, 
with the seven-dimensional theory being a non-compact
gauged $SO(2,2)$ supergravity.  A further field truncation is then 
required in six dimensions.  However, for our present purposes this
embedding is not useful, because, crucially, the Kaluza-Klein vector of
the circle reduction from seven to six dimensions is in fact set to zero. 
This means the seven-dimensional lift of the $S^2$ Pauli reduction would
merely give an $S^1\times S^2$ reduction, with the consistency of the 
$S^2$ reduction remaining unexplained. Upon further investigation we find that,
at least at the bosonic level, there is in fact a different way to 
embed the Salam-Sezgin theory into the seven-dimensional $SO(2,2)$
gauged supergravity, in which the Kaluza-Klein vector plays an active role.
In this new embedding it supplies the necessary twist of the $S^1$ fibres
so that the lift of the $S^2$ Pauli reduction now becomes an $SU(2)$ DeWitt
reduction from seven dimensions.  Furthermore, we find that in this case
the necessary additional truncation of fields in six dimensions is 
compatible with the structure of the Hopf fibration, and so we are able to
reconstruct the $S^2$ Pauli reduction that was obtained in \cite{gibpopss} 
as a DeWitt reduction from seven dimensions, and thus now with an understanding
of why it works.

\section{$S^1$ Reduction of Minimal $D=6$ Supergravity}\label{KKredsec}

   Our starting point is the bosonic sector of minimal six-dimensional
supergravity, for which the equations of motion are
\be
\hat R_{MN} = \ft 18 \hat H_{M}{}^{PQ} \hat H_{NPQ}, \qquad d \hat H_\3 = 0, 
\qquad \hat * \hat H_\3 = \hat H_\3\,.  \label{6Deom}
\ee
Note that since $\hat H_\3=d\hat B_\2$ is self dual, one cannot write a
six-dimensionally covariant Lagrangian for the theory.  We then
perform a Kaluza-Klein $S^1$ reduction, using the standard ansatz
\bea
d\hat s_6^2 &=& e^{2\bar\alpha\phi}\,  d\bar s_5^2 +  
g^{-2}\, e^{2\bar\beta\phi}\, 
(d\tau + g A)^2\,,  \label{mets1} \\
\hat B_\2 &=& B_\2 + g^{-1}\, B_\1\,\wedge d \tau\,, \nn
\eea
where we choose $\bar\alpha^2=1/24$ and $\bar\beta=-3\bar\alpha$ in order
to get the five-dimensional theory in the Einstein frame, and with the 
canonical normalisation for the dilaton field $\phi$. The constant
$g$ that we have introduced here has the dimensions of (length)$^{-1}$,
and serves the purpose of scaling the dimensionless coordinate
$\tau$ that parameterises the circle of the 6th dimension, so as
to give a coordinate $z=g^{-1}\, \tau$ with the dimensions of length.   
Note that we are using hats to denote fields in the original 
six-dimensional theory.  We place bars on five-dimensional quantities in
cases where it is appropriate to distinguish them from six-dimensional
quantities.

The ansatz for
$\hat B_\2$ implies that we shall have
\be
\hat H_\3 = H_\3 + g^{-1}\, H_\2 \wedge (d \tau + g A)\,,\label{hatHH}
\ee
where
\be
H_\3 = d B_\2 - d B_\1 \wedge A\,,\qquad H_\2=dB_\1\,.
\ee
Since the 6-dual of $\hat H_\3$ is given by
\be
\hat * \hat H_\3 = - g^{-1}\, e^{-4\bar \alpha \phi } \, \bar * H_\3 \, 
 \wedge (d\tau + g A) +  e^{4\bar \alpha \phi} \, \bar * H_\2\,,
\ee
the six-dimensional self-duality condition $\hat* \hat H_\3=\hat H_\3$ 
implies the five-dimensional condition
\be
H_\3 = e^{4\bar\alpha\phi}\, \bar * {H_\2} \,, \label{h3h2} 
\ee
so the reduction (\ref{hatHH}) becomes
\be
\hat H_\3 = e^{ 4 \bar\alpha \phi}\, \bar * {H_\2} +  
    g^{-1}\, H_\2\,\wedge(d\tau + g A). \label{h3s1}
\ee
for the case of the self-dual $\hat H_\3$.

   The five-dimensional equations of motion resulting from substituting the 
reduction ans\"atze into the six-dimensional equations (\ref{6Deom})
 can be derived from the Lagrangian
\be
{\cal L}_5= \bar R\,  \bar*\oneone-\ft12 \bar *d \phi \wedge d \phi-
 \ft12 e^{-8 \bar\alpha \phi} \, \bar *F_\2 \wedge F_\2 - 
 \ft12  e^{4 \bar \alpha \phi}\,  \bar *H_\2 \wedge H_\2 - 
\ft12 H_\2 \wedge H_\2 \wedge A_\1 \,. \label{lag5}
\ee
This Lagrangian describes the bosonic sector of five-dimensional minimal
supergravity coupled to one vector multiplet.  The truncation to
pure minimal supergravity is then achieved by setting
\be
\phi=0\,,\qquad B_\1 = \sqrt{2}\, A_\1\,,\label{truncation}
\ee
which can easily be seen to be consistent with the equations of motion.  If
we define $\wtd A_\1 = \sqrt3\, A_\1$ so that the remaining gauge field
has a canonical normalisation, the Lagrangian (\ref{lag5}) reduces to
that for the bosonic sector of pure minimal supergravity:
\be
{\cal L}_5= \bar R\,  \bar*\oneone -
 \ft12 \bar *\wtd F_\2 \wedge \wtd F_\2 -
\ft1{3\sqrt3}\,  \wtd F _\2 \wedge \wtd F_\2 \wedge \wtd A_\1 \,. 
\label{lag5min}
\ee

\section{$SU(2)$ DeWitt Reduction from $D=6$ to $D=3$}\label{6to3DeWitt}

\subsection{Description as an $SU(2)$ group manifold reduction}
    \label{su2subsec}\label{6to3sec}

   The $SU(2)$ group manifold DeWitt reduction of minimal six-dimensional
supergravity was constructed in \cite{lupose}.  The reduction ansatz
for the metric and the self-dual 3-form are given by
\bea
d\hat s^2_6 &=& e^{2\alpha \vp} d s^2_3 + 
g^{-2}\, e^{2\beta \vp} \, {\wtd T}_{i j}\, \
\nu^i \, \nu^j \,, \label{metsu2}\\
\hat H_\3 &=& m g^{-3} \Omega_\3 + m e^{4\alpha \vp}\, \ep_\3
+ \ft12 g^{-2} \,\vep_{i j k}\, B^i 
\wedge \nu^j \wedge \nu^k - 
g^{-1} e^{\fft{4\alpha \vp} 3}\,\wtd T_{i j} \,{* B}^i\wedge\,\nu^j \,,
\label{H3ans}
\eea
where the constants $\alpha$ and $\beta$ are taken to be given by
$\alpha^2=3/8$ and $\beta=-\alpha/3$ in order to obtain the 
three-dimensional theory in Einstein frame with the
canonical normalisation for the
dilaton $\vp$.  The unimodular matrix $\wtd T_{ij}$ parameterises the
remaining scalar fields of the three-dimensional theory, and the
1-forms 
\be
\nu^i=\sigma^i- g A^i\,,\label{nui}
\ee
 are written in terms of the left-invariant
1-forms $\sigma^i$ of $SU(2)$ and the $SU(2)$ Yang-Mills potentials $A^i$,
and 
\be
\Omega_\3\equiv \nu^1\wedge\nu^2\wedge\nu^3\,, \label{omeg3}
\ee
$\ep_\3$ is the volume form of the three-dimensional spacetime, and 
$B^i$ denotes an $SU(2)$ triplet of 1-form fields.

The $\sigma^i$, which can be expressed in terms of Euler angles 
$(\psi,\theta,\tau)$ as
\be
\sigma_1 = \cos\psi\, d\theta + \sin\psi\, \sin\theta\, d\tau\,,\quad
\sigma_2 = -\sin\psi\, d\theta + \cos\psi\, \sin\theta\, d\tau\,,\quad
\sigma_3 = d\psi + \cos\theta\, d\tau\,, \label{sigmai}
\ee
obey the relations
\be
d\sigma_i= -\ft12 \vep_{ijk}\, \sigma_j\wedge \sigma_k\,.
\ee
We also have
\be
D\nu^i = -\ft12\vep_{ijk}\, \nu^j\wedge \nu^k -g F^i\,,
\ee
where
\be
F^i= dA^i + \ft12 g\vep_{ijk}\, A^j\wedge A^k\,,\qquad D\nu^i\equiv d\nu^i + g
   \vep_{ijk}\, A^j\wedge \nu^k\,.
\ee
Note that the equation $d\hat H_\3=0$ implies that
\be
DB^i - m F^i + g e^{\ft43\alpha\vp}\, \wtd T_{ij}\, {*B}^j=0\,,\label{DB}
\ee
where $DB^i=dB^i + g \vep_{ijk}\, A^j\wedge B^k$.

  The equations of motion for the three-dimensional theory, obtained by
substituting (\ref{metsu2}) and (\ref{H3ans}) into (\ref{6Deom}), can
be derived from a Lagrangian whose precise form can be found in \cite{lupose}.

\subsection{$SU(2)$ as a Hopf fibration} \label{pauliredsec}

  Following \cite{paulired}, we may now rewrite the $SU(2)$ DeWitt reduction
of subsection \ref{su2subsec} in a form where $SU(2)$ is viewed as a 
$U(1)$ Hopf fibration over $S^2$.   Thus we describe the unit $S^2$ 
as the surface $\mu_1^2 + \mu_2^2 +\mu_3^2=1$ in $\R^3$, where the three
Cartesian coordinates $\mu_i$ are parameterised in terms of the $\theta$ and
$\psi$ Euler angles introduced in (\ref{sigmai}) by
\be
\mu_1 = \sin\psi\, \sin\theta\,,\quad
\mu_2 = \cos\psi\, \sin\theta\,,\quad
\mu_3 = \cos\theta\,.
\ee
The 1-forms $\nu^i$ defined in (\ref{nui}) can then be written as 
\cite{paulired} 
\bea
\nu^i &=& \sigma^i -g \, A^i = - \vep_{i j k}\, \mu^j\, D \mu^k + 
\mu^i\, \sigma\,, \nn\\
\sigma &\equiv& d\tau + \cos\theta\, d\psi - g \, \mu^i\, A^i, \label{sigma}
\eea
where the covariant derivative is defined as 
$D\mu^i \equiv d\mu^i + g \,\vep_{i j k}\, A^j\, \mu^k\,$. 

   The metric reduction ansatz (\ref{metsu2}) can now be seen to be
given by \cite{paulired}
\be
d\hat s_6^2 = e^{2\alpha\vp}\, ds_3^2 + g^{-2}\, 
 e^{2\beta\vp}\, \wtd\Delta^{-1}\, 
\wtd T^{-1}_{ij}\, D\mu^i\, D\mu^j + g^{-2}\, e^{2\beta\vp}\, \wtd\Delta\, 
(d\tau + g A)^2\,,\label{hopfmetric}
\ee
where
\be
A = g^{-1}\, \cos\theta\, d\psi - \mu^i A^i - 
  g^{-1}\, \wtd\Delta^{-1}\, \wtd T_{ij}\,  
\vep_{ik\ell}\, \mu^j \mu^k\, D\mu^\ell\,,
\qquad
 \wtd\Delta = \wtd T_{ij}\, \mu^i \mu^j\,. \label{Aminsugra}
\ee

    After some algebra, we find we can write
\bea
\nu^i &=& \mu^i\, (d\tau + g A) -\wtd\Delta^{-1}\, \wtd T_{jk}\, 
            \vep_{ij\ell}\, \mu^k\, D\mu^\ell\,,\nn\\
\ft12\vep_{ijk}\, \nu^j\wedge \nu^k &=&
  (d\tau+g A)\wedge D\mu^i + \wtd\Delta^{-1}\, \wtd T_{ij}\, \mu^j\, \omega_\2
\,,\nn\\
\Omega_\3 &=& \ft16 \vep_{ijk}\nu^i\wedge\nu^j\wedge\nu^k= 
   (d\tau+gA) \wedge \omega_\2\,,\label{Om3etc}
\eea
where
\be
\omega_\2 = \ft12 \vep_{i j k}\,\mu^i\, D\mu^j\,\wedge\, D\mu^k\,.
\ee
 It then follows that the reduction ansatz (\ref{H3ans}) for the 
self-dual 3-form
is given by
\bea
\hat H_\3 &=& (d\tau+ g A)\wedge \Big[m\,g^{-3}\,
\omega_\2 - g^{-2}\, B^i\wedge D\mu^i -
  g^{-1}\, e^{\ft43\alpha\vp}\, \wtd T_{ij} \, \mu^i\, {*B}^j \Big]
\label{h3su2} \\
&& +m e^{4\alpha\vp}\, \ep_\3 + g^{-2}\, \wtd\Delta^{-1} \,
\wtd T_{ij}\, \mu^i B^j\wedge \omega_\2 
 +g^{-1} e^{\ft43\alpha\vp}\, \wtd\Delta^{-1}\, 
\vep_{jkm}\, \mu^\ell\, \wtd T_{ij}\,
  \wtd T_{k\ell}\,   {*B}^i\wedge D\mu^m\,.\nn
\eea

   With these preliminaries, we are now ready to re-interpret the
DeWitt $SU(2)$ group manifold reduction of the minimal six-dimensional
supergravity as a Pauli $S^2$ reduction from five dimensions.  To do this, 
we compare the expressions (\ref{mets1}) and (\ref{h3s1}) for the
$S^1$ reduction with the corresponding expressions (\ref{hopfmetric}) and
(\ref{h3su2}) for the $SU(2)$ reduction expressed in the notation of the Hopf
fibration.  Thus from the comparison of the metrics we find
\bea
d\bar s_5^2 &=& e^{2\alpha\vp-2\bar\alpha\phi}\, ds_3^2 + 
     g^{-2}\, e^{-\ft23\alpha\vp - 2\bar\alpha\phi}\,\wtd\Delta^{-1}\,
\wtd T^{-1}_{ij}\, D\mu^i D\mu^j\,,\\
e^{-6\bar\alpha\phi} &=& e^{-\ft23\alpha\vp}\, \wtd\Delta\,,
\eea
and from the comparison of the reduction ans\"atze for the six-dimensional
self-dual 3-form $\hat H_\3$ we find
\bea
H_\2 &=& m g^{-2}\, \omega_\2 - g^{-1}\, B^i\wedge D\mu^i -
  e^{\ft43\alpha\vp}\, \wtd T_{ij}\, \mu^i\, {*B}^j\,,\\
e^{4\bar\alpha\phi} \, {\bar * H_\2} &=& 
   m e^{4\alpha\vp}\,\ep_\3 + g^{-2} \, \wtd\Delta^{-1} \,
\wtd T_{ij}\, \mu^i B^j\wedge \omega_\2 
 +g^{-1} e^{\ft43\alpha\vp}\, \wtd\Delta^{-1}\, 
\vep_{jkm}\, \mu^\ell\, \wtd T_{ij}\,
  \wtd T_{k\ell}\,  {*B}^i\wedge D\mu^m\,.\nn
\eea

    Following \cite{paulired}, we now define the three-dimensional
scalar fields
\be
T_{ij}= Y^{\ft13}\, \wtd T_{ij}\,,\qquad Y= e^{4\alpha\vp}\,,\label{Tijdef}
\ee
in terms of which the Pauli reduction ans\"atze for the five-dimensional
metric $d\bar s_5^2$ and fields $\phi$, $A$ and $H_\2=dB_\1$ become
\bea
d\bar s_5^2 &=& Y^{\ft13}\, \Delta^{\ft13}\, ds_3^2 + g^{-2}\, Y^{\ft13}\,
\Delta^{-\ft23}\, T^{-1}_{ij}\, D\mu^i D\mu^j\,,\nn\\
e^{6\bar\alpha \phi} &=& Y^{\ft12}\, \Delta^{-1}\,,\nn\\
A &=& g^{-1}\, \cos\theta d\psi -\mu^i\, A^i - g^{-1}\, \Delta^{-1}\,
    T_{ij}\, \vep_{ik\ell}\, \mu^j \mu^k \, D\mu^\ell\,,\nn\\
H_\2 &=&  m g^{-2}\, \omega_\2 - g^{-1}\, B^i\wedge D\mu^i -
              T_{ij}\, \mu^i\, {*B}^j\,,\label{5to3red}
\eea
where $\Delta= T_{ij}\, \mu^i \mu^j = Y^{\ft13}\, \wtd\Delta$.

   Making use of the equation (\ref{DB}), we can see that $H_\2=dB_\1$ 
given in  (\ref{5to3red}) can be written as
\be
H_\2= m g^{-2}\, \omega_\2 -mg^{-1}\, \mu^i F^i + g^{-1}\, d(\mu^i B^i)\,,
\ee
and hence $B_\1$ can be written explicitly as
\be
B_\1 = mg^{-2}\, \cos\theta\, d\psi - m g^{-1}\, \mu^i A^i + 
   g^{-1}\, \mu^i B^i\,.\label{B1red}
\ee

\section{Pauli reduction of $5D$ Minimal Supergravity?}

   In section \ref{pauliredsec} we constructed the consistent $S^2$ Pauli
reduction of the bosonic sector of the five-dimensional supergravity that
is obtained by means of the $S^1$ Kaluza-Klein reduction of minimal
six-dimensional supergravity.  As we showed in section \ref{KKredsec}, the
five-dimensional theory can be truncated to give the bosonic sector of
pure minimal five-dimensional supergravity by imposing the conditions 
(\ref{truncation}) on the five-dimensional fields.  In this section, we
address the question of whether we can consistently impose this 
truncation on the three-dimensional fields in the Pauli reduction in 
section \ref{pauliredsec}, thereby obtaining a consistent $S^2$ Pauli 
reduction of five-dimensional minimal supergravity.

  From the Pauli reduction ansatz for $\phi$ given in (\ref{5to3red}), we
see that setting $\phi=0$ requires imposing
\be
Y^{\ft12} = \Delta = T_{ij}\, \mu^i \mu^j\,.
\ee
Since $Y$ and $T_{ij}$ are three-dimensional fields, which cannot depend on the
$S^2$ coordinates $\mu^i$, it follows that we must have
\be
T_{ij} = f\delta_{ij}\,,
\ee
where $f$ is a function only of the three-dimensional fields.  Taking the
determinant of $T_{ij}$, and noting from (\ref{Tijdef}) that it must equal
$Y$, we then conclude that $f^2=f^3$ and hence $f=1$, so $T_{ij}=\delta_{ij}$.
 From the reduction ans\"atze for $A$ and $B_\1$ given in (\ref{5to3red})
and (\ref{B1red}), we then conclude that making the truncation $B_\1=\sqrt 2 A$
in (\ref{truncation}) implies
\be
 m= \sqrt2\, g\,,\qquad B^i=0\,.
\ee
 Finally, from (\ref{DB}) we see that $B^i=0$ implies that $F^i=0$, and
so $A^i$ is pure gauge.

   The conclusion from the above discussion is that one cannot truncate the
consistent $S^2$ Pauli reduction of the full five-dimensional theory to
give a consistent Pauli reduction of the five-dimensional minimal 
supergravity theory.

\section{Salam-Sezgin Theory by $S^1$ reduction from $D=7$}

   First, we need to see how to obtain the $D=6$ Salam-Sezgin theory from an
$S^1$ reduction of a $D=7$ theory.  In \cite{cvluposatr}, the details
of the $S^3$ reduction from $D=10$, giving ${\cal N}=4$ supersymmetric 
gauged $SO(4)$ sugra in $D=7$ were given; it was obtained as a limit of
the gauged $SO(5)$ supergravity that comes from $S^4$ reduction from 
$D=11$.
The $SO(4) \rightarrow SO(2,2)$ replacement was then discussed in
\cite{cvgiposs}, where it was then shown how Salam-Sezgin 
could be obtained via an
$S^1$ reduction of gauged $SO(2,2)$ ${\cal N}=2$ supergravity by $S^1$
reduction followed by a consistent truncation.  Prior to the final
consistent truncation in $D=6$, the bosonic Lagrangian is given by
eqn (30) of \cite{cvgiposs}.  We can straightforwardly make a
truncation of all $SO(2,2)$ fields to those that are singlets under the
$U(1)\times U(1)$ maximal subgroup.  The truncated seven-dimensional
bosonic Lagrangian from which the reduction to six dimensions can be obtained
is
\bea
{\cal L}_7 &=& \hat R \,\hat *\oneone - 
\ft5{16} \hat \Phi ^{-2} \, \hat* d \hat \Phi \wedge d \hat \Phi
-\hat\Phi^{-1/2}\,\, ({*\hat F_\2^{12}}\wedge \hat F_\2^{12} +
  {*\hat F_\2^{34}}\wedge \hat F_\2^{34})
- \ft12 \hat \Phi^{-1} \, \hat * \hat H_\3 \wedge \hat H_\3 \nn\\
&& -4 g^2 \, \hat \Phi^{\ft12} \, \hat * \oneone +{\cal L}_{7,CS},
\label{lag7trunc}
\eea
where the $SO(2,2)$ gauge potentials $\hat A_\1^{\alpha\beta}$ have been
truncated to just the abelian subsector $\hat A_\1^{12}$ and $\hat A_\1^{34}$,
and $\hat H_\3= d\hat B_\2 + \hat F_\2^{12}\wedge \hat A_\1^{34} +
 \hat F_\2^{34}\wedge \hat A_\1^{12}$.
Performing a Kaluza-Klein circle reduction in the usual way,
by means of the ans\"atze\footnote{We shall use a bar to 
denote six-dimensional quantities, such as the metric and $\bar B_\2$,
where we may need to distinguish
them later from four-dimensional quantities, which will be unbarred.  
In cases where there is no possibility of confusion with four-dimensional
quantities, we shall omit the bar, as, for example, 
in the six-dimensional gauge fields $A_\1^{12}$ and $A_\1^{34}$.}
\bea
d\hat s_7^2 &=& e^{2\alpha\varphi}\, d\bar s_6^2 + 
                e^{-8\alpha\varphi}\, (dz+ \cA_\1)^2\,,\nn\\
\hat B_\2 &=& \bar B_\2 + \bar B_\1\wedge dz\,,\qquad 
\hat A_\1^{\alpha\beta}= A_\1^{\alpha\beta}\,,\qquad
\hat\Phi=\Phi\,,\label{7to6red}
\eea
where $\alpha=1/(2\sqrt{10})$, gives the six-dimensional 
Lagrangian\footnote{There were some typographical errors in 
\cite{cvluposatr}, which we have corrected here, 
relating to the coefficient of the kinetic term for the scalar field $\Phi$,
and also the form of the Chern-Simons term, whose variation is given
by (\ref{deltaLCS}).}
\bea
{\cal L}_6 &=& \bar R {\bar *\oneone} - 
   \ft{5}{16} \Phi^{-2}\, {\bar *d\Phi}\wedge d\Phi
- \ft12{\bar *d\varphi}\wedge d\varphi - \ft12 e^{-10\alpha\varphi} \,
{\bar *{\cal F}}_\2\wedge {\cal F}_\2
-\ft12 \Phi^{-1}\, e^{6\alpha\varphi}\, {\bar * \bar H_\2}\wedge \bar H_\2\nn\\
&&
 -\Phi^{-1/2}\, e^{-2\alpha\varphi}\, ({\bar *F_\2^{12}}\wedge F_\2^{12} +
  {\bar *F_\2^{34}}\wedge F_\2^{34})  
 -\ft12 \Phi^{-1}\, e^{-4\alpha\varphi}\, {\bar *\bar H_\3}\wedge \bar H_\3\nn\\
&&-4 g^2\, \Phi^{1/2}\, e^{2\alpha\varphi}\, {\bar *\oneone} 
   + {\cal L}_{CS}\,,
\label{D6abelian}
\eea
where
\bea
\bar H_\3&=& d\bar B_\2 - d\bar B_\1\wedge {\cal A}_\1 + 
   F_\2^{12}\wedge A_\1^{34} +
   F_\2^{34}\wedge A_\1^{12}\,,\nn\\
\bar H_\2 &=& d\bar B_\1\,,\qquad {\cal F}_\2 = d{\cal A}_\1\,,\label{H3H2}
\eea
and $\bar *$ denotes the six-dimensional Hodge dual in the metric 
$d\bar s_6^2$.
The term ${\cal L}_{CS}$ is a Chern-Simons term, whose variation is given,
up to a certain overall normalisation constant $c$, by
\be
\delta {\cal L}_{CS}= c (F_\2^{\beta\gamma}\wedge F_\2^{\gamma\delta}\wedge
F_\2^{\delta\alpha}- \ft14 F_\2^{\gamma\delta}\wedge F_\2^{\delta\gamma}\wedge
  F_\2^{\beta\alpha})\wedge \delta A_\1^{\alpha\beta}\,.\label{deltaLCS}
\ee
An important point for the consistency of the truncation that gives the
Salam-Sezgin theory is that this Chern-Simons contribution vanishes if
one sets $A_\1^{12}= \pm A_\1^{34}$.

Note that we have, for reasons of presentational simplicity, 
omitted the axions that would come 
from the reduction of the seven-dimensional gauge fields 
$\hat A_\1^{12}$ and $\hat A_\1^{34}$.  Setting them to zero 
would not in general be a
consistent truncation, but it is consistent to do so in either of
the two further truncations that we shall be considering below, namely
either setting $A_\1^{12}= - A_\1^{34}$ with $\cA_\1=B_\1=0$, or else 
setting $A_\1^{12}=A_\1^{34}=0$ with $\cA_\1=-B_\1$.   

  It is convenient to re-parameterise the 
scalars in terms of the two fields $\phi$ and $\psi$, where
\be
\Phi= e^{\ft25 \psi -\ft45\phi}\,,\qquad 20\alpha\varphi= -2\psi-\phi\,. 
\label{repar}
\ee
In terms of these, the Lagrangian becomes
\bea
{\cal L}_6 &=& \bar R {\bar *\oneone} -\ft14 {\bar *d\phi}\wedge d\phi -
  \ft14 {\bar *d\psi}\wedge d\psi - \ft12 e^{\ft12\phi+\psi} \,
{\bar *{\cal F}}_\2\wedge {\cal F}_\2 -
\ft12 e^{\ft12\phi-\psi}\, {\bar *\bar H_\2}\wedge \bar H_\2\nn\\
&& 
 - e^{\ft12\phi}\, ({\bar *F_\2^{12}}\wedge F_\2^{12} +
  {\bar *F_\2^{34}}\wedge F_\2^{34}) 
-\ft12 e^{\phi}\, {\bar *\bar H_\3}\wedge \bar H_\3 
-4 g^2\, e^{-\ft12\phi}\, {\bar *\oneone} + {\cal L}_{CS}\,.\label{D6pre}
\eea

\subsection{Truncations to Salam-Sezgin theory}

    It was shown in \cite{cvgiposs} that the Salam-Sezgin theory 
could be obtained by making a further, consistent, truncation
of the six-dimensional supergravity whose relevant bosonic sector is
described by (\ref{D6pre}).  Namely, one now sets
\be
\cA_\1=0\,,\qquad \bar B_\1=0\,,\qquad 
A_\1^{12}= - A_\1^{34}\equiv \ft12 A_\1\,,
\qquad\psi=0\,,\label{trunc1}
\ee
leading to the Salam-Sezgin bosonic Lagrangian
\be
{\cal L}_{SS} =  \bar R \, {\bar *\oneone} -\ft14 {\bar *d\phi}\wedge d\phi -
  \ft12 e^{\ft12\phi}\, {\bar *F_\2}\wedge F_\2 -
   \ft12 e^{\phi}\, {\bar *\bar H_\3}\wedge \bar H_\3 -
           4 g^2\, e^{-\ft12\phi}\, {\bar *\oneone}
\,.\label{salsezlag}
\ee
(Recall that, as already remarked, the setting to zero of the axions coming 
from the reduction of the $U(1)\times U(1)$ Yang-Mills potentials is
consistent, once the truncation (\ref{trunc1}) is performed.  Furthermore,
the Chern-Simons contribution vanishes under this truncation.)

   This construction, and its extension to include the fermionic sector also,
was studied in detail in \cite{cvgiposs}.  It gives a 
consistent embedding of the Salam-Sezgin theory in a seven-dimensional 
gauged supergravity, which in turn can be obtained as a consistent reduction
of ten-dimensional supergravity.  However, it does not provide us with a 
way to understand the occurrence of the consistent Pauli $S^2$ reduction 
\cite{gibpopss} of the Salam-Sezgin theory itself.  The understanding
of a consistent Pauli $S^2$ reduction from $D$ dimensions by first 
considering a (trivially)
consistent DeWitt $SU(2)$ group manifold reduction from $(D+1)$ dimensions
depended upon the $S^2$ reduction becoming an $SU(2)$ reduction when lifted
to the higher dimension.  This depends upon the Kaluza-Klein vector of the
$(D+1)\rightarrow D$ reduction providing the necessary non-trivial monopole
background that twists the $S^1$ into a Hopf fibration over the $S^2$,
becoming the $SU(2)$ group manifold.  In the construction in 
\cite{cvgiposs}, however, the Kaluza-Klein vector is actually
set to zero, as in (\ref{trunc1}), and so the lift to $D=7$ of the 
consistent Pauli $S^2$ reduction of the Salam-Sezgin theory that was
found in \cite{gibpopss} will be an $S^1\times S^2$ reduction rather than
an $SU(2)$ group-manifold reduction.

   If we are to find an explanation of the consistency of the $S^2$ Pauli
reduction of the Salam-Sezgin theory in terms of a Hopf reduction of an
$SU(2)$ group manifold reduction from $D=7$, we must find a
different embedding of the Salam-Sezgin theory into $D=7$, in which the 
Kaluza-Klein vector plays the role of supplying the necessary monopole 
twist.  At least at the bosonic level, the existence of such an alternative
reduction can be seen by looking again at the six-dimensional 
Lagrangian (\ref{D6pre}).  Now, we set
\be
\cA_\1=- \bar B_\1 \equiv \fft1{\sqrt2}\, A_\1\,,\qquad
A_\1^{12}=A_\1^{34}=0\,,\qquad \psi=0\,.\label{trunc2}
\ee
It is straightforward to check that this is indeed a consistent truncation,
and that it yields the same Salam-Sezgin bosonic Lagrangian (\ref{salsezlag})
that we saw previously.  (Again, the setting to zero of the 
axions coming from the reduction of the $U(1)\times U(1)$ gauge potentials
is indeed consistent, under this truncation, and the Chern-Simons term
again gives zero contribution.)

\section{DeWitt and Hopf Reduction from $D=7$}

   Having seen that we can indeed obtain the bosonic Salam-Sezgin theory
from seven dimensions in a circle reduction where the Kaluza-Klein vector
is active, we now turn to the question of whether we can use this to obtain the
consistent Pauli $S^2$ reduction of Salam-Sezgin via a Hopf reduction of 
the seven-dimensional theory.  The calculations here will be closely 
analogous to those that we carried out in section \ref{6to3DeWitt}.  
Accordingly, we begin by considering the $SU(2)$ group manifold DeWitt
reduction from $D=7$, with the standard metric ansatz\footnote{Note that the
vacuum solution $S^3\times$(Minkowski)$_4$ of the theory described by
(\ref{lag7trunc}) has $d\hat s_7^2= dx^\mu dx_\mu + g^{-2}\, d\Omega_3^2$
together with $\hat H_\3=\pm 2g\, \tilde\Omega_\3$ and $\hat\Phi=1$, 
where $d\Omega_3^2$
is the metric on the unit $S^3$ and $\tilde\Omega_\3$ is its volume form.
(Note that $\tilde\Omega_\3=\ft18 \Omega_\3$ in the vacuum, where
$\Omega_\3$ was defined earlier in eqn (\ref{Om3etc}).)}
\be
d\hat s^2_7 = e^{2\alpha' \vp'} d s^2_4 +
\ft14 g^{-2}\, e^{-\ft{4\alpha'}3 \vp'} \, T_{i j}\, \
\nu^i \, \nu^j \,, \label{74metsu2}\\
\ee
where $\nu^i= \sigma^i - g A^i$, with $\sigma^i$ being the 
left-invariant 1-forms of $SU(2)$, as described in section \ref{6to3sec},
and $\alpha'^2=\ft3{20}$. The matrix of scalar fields $T_{ij}$ is 
unimodular. 

   Following the
same strategy as we did in section \ref{6to3DeWitt}, we now write this in terms 
of the Hopf fibration, which here will take the form
\be
d\hat s_7^2 = e^{2\alpha' \vp'}\, ds_4^2 + \ft14 g^{-2}\,
 e^{-\ft{4\alpha'}3 \vp'}\, \Delta^{-1}\,
T^{-1}_{ij}\, D\mu^i\, D\mu^j + e^{-\ft{4\alpha'}3 \vp'}\, \Delta\,
(dz+ {\cal A}_\1)^2\,,\label{met7to4}
\ee
where
\be
{\cal A}_\1 = \ft12 g^{-1}\, \cos\theta\, d\psi - \ft12 \mu^i A^i -
  \ft12 g^{-1}\, \Delta^{-1}\, T_{ij}\,
\vep_{ik\ell}\, \mu^j \mu^k\, D\mu^\ell\,,
\qquad
 \Delta = T_{ij}\, \mu^i \mu^j\,. \label{calA}
\ee
Comparing (\ref{met7to4}) with the $S^1$ reduction of the metric in
(\ref{7to6red}), we see we must have 
\be
e^{-8 \alpha  \vp} = e^{-\ft{4\alpha'}3 \vp'}\, \Delta \,, \label{phdel}
\ee
together with
\be
d\bar s^2_6 = e^{-2 \alpha \vp + 2\alpha' \vp'} d s^2_4
+ \ft14 g^{-2}\, 
 e^{-2 \alpha \vp - \ft{4\alpha'}3 \vp'}\, \Delta^{-1}\,
T^{-1}_{ij}\, D\mu^i\, D\mu^j \,. \label{met6to4}
\ee
A straightforward calculation from (\ref{calA}) shows that
\be
{\cal F}_\2 = -\ft12 g^{-1} \, U \Delta ^{-2} \,\omega_\2
+ \ft12 g^{-1} \Delta ^{-2} \varepsilon_{ijk} \, D\mu^i 
\wedge D T_{j\ell}\,T_{km}\, \mu^{\ell} \mu^m
- \ft12 \Delta ^{-1} T_{ij} \mu^i \, F^j \,, \label{calF}
\ee
where $U = 2 T_{ik} \, T_{kj} \, \mu^i \mu^j - \Delta \, T_{ii}$.

   The general DeWitt reduction of the 3-form in seven dimensions is given by
\be
\hat H_\3 = m g^{-3}\, \Omega_\3 + g^{-2}\, 
\ft12 \varepsilon_{i j k} \,B^i \wedge \nu^j \wedge \nu^k
+ g^{-1} \, C^i \wedge \nu^i + H_\3,
\ee
where $\Omega _\3$ was defined in (\ref{omeg3}), $B^i$ and $C^i$ 
are $SU(2)$ triplets of four-dimensional 1-form and 2-form fields respectively,
and $H_\3$ is a four-dimensional 3-form. 
Using the relations 
\bea
\nu^i &=& 2 g \mu^i\, (dz+ {\cal A}_\1) - \Delta^{-1}\, T_{jk}\, 
            \vep_{ij\ell}\, \mu^k\, D\mu^\ell\,,\nn\\
\ft12\vep_{ijk}\, \nu^j\wedge \nu^k &=&
 2 g (dz+ {\cal A}_\1) \wedge D\mu^i + 
  \Delta^{-1}\, T_{ij}\, \mu^j\, \omega_\2
\,,\nn\\
\Omega_\3 &=& \ft16 \vep_{ijk}\nu^i\wedge\nu^j\wedge\nu^k= 
   2 g (dz+ {\cal A}_\1) \wedge \omega_\2\,,
\eea
and comparing with the $S^1$ reduction for $\hat B_2$ in (\ref{7to6red}), and
the six-dimensional definitions (\ref{H3H2}) (recall we are setting the
seven-dimensional $U(1)\times U(1)$ gauge fields to zero in this 
truncation), we find that
\bea
\bar H_\2 &=& 2 m g^{-2}\,\omega_\2 - 2 g^{-1} B^i \wedge D \mu^i + 2 
\mu^i C^i, \nn\\
\bar H_\3 &=& g^{-2}\, \Delta^{-1}\, T_{ij}\, \mu^j\, B^i \wedge \omega_\2
-g^{-1}\, \Delta^{-1}\, T_{jk}\, \varepsilon_{i j \ell}\, \mu^k\, 
C^i \wedge D \mu^\ell + H_\3. \label{h2h3}
\eea

  With these reduction ans\"atze for the metric and the 3-form, we are thus
able to give the complete DeWitt $SU(2)$ reduction of the seven-dimensional
theory described by the Lagrangian (\ref{lag7trunc}) (with $\hat A_\1^{12}=
\hat A_\1^{34}=0$ and hence the Chern-Simons term making no contribution 
either).
The reduction is expressed in the form of the Hopf fibration, allowing us
to make contact with the $S^1$ reduction we discussed earlier.  Specifically,
we now wish to impose the further truncation of the six-dimensional fields
given in (\ref{trunc2}), which takes us in six dimensions to the Salam-Sezgin 
theory.  

   As we saw in a similar discussion in section \ref{6to3DeWitt}, 
it is not guaranteed that imposing this truncation will be compatible with
retaining the desired fields in the DeWitt-Hopf reduction.  The key question
here is whether imposing the truncation $\cF_\2= -H_\2$ is compatible with
the expressions in (\ref{calF}) and (\ref{h2h3}) for these fields.  
Equating the two expressions, we find
\be
U =4 m \, g^{-1} \,\Delta ^{2} \,, \qquad
B^i = -\ft14 \Delta ^{-2} \varepsilon_{ijk} \,D T_{j\ell}\,T_{km}\, 
\mu^{\ell} \mu^m \,, \qquad
C^i = \ft14 \Delta ^{-1} T_{ij}\, F^j\,. \label{h=-f}
\ee
Following similar arguments to those we presented in section \ref{6to3DeWitt}, 
we conclude here that performing the truncation is consistent, provided
we have
\be
T_{ij} = \delta_{ij} \,, \qquad m = -\ft14 g  \,, \qquad B^i = 0 \,, 
\qquad C^i = \ft14 F^i \,.
\ee

   Going back to (\ref{phdel}), we now find $\alpha'\varphi'= 6 \alpha\varphi$,
and hence from (\ref{repar}), with the truncation $\psi=0$ that we have made,
we find the Pauli metric reduction ansatz (\ref{met6to4}) reduces to
\be
d\bar s_6^2 = e^{-\ft12\phi}\, ds_4^2 + \ft14 g^{-2}\, e^{\ft12\phi}\, 
     D\mu^i D\mu^i\,.\label{paulimet}
\ee
The field strengths reduce according to the ans\"atze 
\bea
F_\2 &=& \sqrt2 \cF_\2=-\sqrt2 \bar H_\2= 
\ft1{\sqrt2} g^{-1} \omega_\2 - \ft1{\sqrt2} \mu^i \,F^i\,, \nn\\
\bar H_\3 &=& H_\3 - 
 \ft14 g^{-1} \varepsilon_{ijk} \,F^i \mu^j \, \wedge D \mu^k\,. \label{sase}
\eea
The scalar field $\phi$ is simply taken to be dependent only on the 
four-dimensional coordinates.

   We may now verify that this reduction scheme does indeed give consistent
four-dimensional equations of motion, as a consequence of the six-dimensional
equations of motion.  First, we check the six-dimensional Bianchi identity
for $\bar H_\3$, which, from (\ref{H3H2}), is
\be
d \bar H_\3 = - \bar H_\2 \wedge {\cal F}_\2 = 
  \ft12 F_\2\wedge F_\2 \,. \label{fh3h2}
\ee
To calculate $d\bar H_\3$, we may employ the useful relations
\be
D F^i = 0\,, \qquad
\ft12 \vep_{i j k}\,D\mu^j\,\wedge\,D\mu^k = \mu^i \, \omega_\2\,,\qquad
D^2 \mu^i =  g \,\vep_{i j k}\, F^j\,\mu^k \,. \label{murel}
\ee
After some algebra, we find $d\bar H_\3$ gives
\be
d\bar H_\3 = dH_\3 - \ft12 g^{-1} \, \mu^i\, F^i \,\wedge \omega_\2
- \ft14 F^i\,\wedge F^i + \ft14 \mu^i \mu^j\, F^i\wedge F^j\,,
\ee
while
\be  
F_\2 \wedge F_\2 = - g^{-1} \, \mu^i\, F^i \,\wedge \omega_\2
 + \ft12 \mu^i \mu^j\, F^i \wedge F^j\,.
\ee
Therefore, all the coordinate dependence on the internal 2-sphere coordinates
$\mu^i$ cancels, and 
(\ref{fh3h2}) leads to the four-dimensional Bianchi identity
\be
d H_\3 = \ft14 F^i\,\wedge F^i\,.
\ee

  To check the other six-dimensional equations of motion, it is useful first
to calculate the six-dimensional duals of the fields $\bar H_\3$ and $F_\2$. 
We find
\bea
{\bar *\bar H_\3} &=& 
  \ft14 g^{-2}\, e^\phi\, {* H_\3}\wedge\omega_\2 
               +\ft14 g^{-1}\, {* F^i}\wedge D\mu^i\,,\nn\\
{\bar *F_\2} &=& 2\sqrt2\, g e^{-\ft32\phi}\, {* \oneone} -
   \fft1{4\sqrt2 g^2}\, e^{\ft12\phi}\,\mu^i\,  {* F^i}\wedge\omega_2\,,
\label{starHF}
\eea
together with ${\bar *d\phi}=  \ft14 g^{-2}\, {* d\phi}\wedge\omega_\2$.
In these expressions, $*$ denotes the Hodge dual in the four-dimensional
metric $ds_4^2$.  

   We find that the six-dimensional equation of motion for the
scalar field $\phi$,
\be
d{\bar *d\phi} + \ft12 e^{\ft12\phi}\, {\bar *F_\2}\wedge F_\2 +
  e^\phi\, {\bar *\bar H_\3}\wedge \bar H_\3 - 
        4 g^2\, e^{-\ft12\phi}\, {\bar *\oneone}=0
\ee
implies, after non-trivial cancellations of the internal coordinate dependence,
the four-dimensional equation
\be
d{*d\phi} + e^{2\phi}\, {*H_\3}\wedge H_\3 +
  \ft14 e^{\phi}\, {* F^i}\wedge F^i=0\,.
\ee

   The six-dimensional equation of motion for $\bar H_\3$, namely 
$d(e^\phi\,{\bar *\bar H_\3})=0$, implies the two four-dimensional equations
\be
d(e^{2\phi}\, {*H_\3})=0\,,\qquad \hbox{and}\quad 
D(e^\phi\, {*F^i}) - e^{2\phi}\, {*H_\3}\wedge F^i=0\,.
\label{bHFieqns}
\ee

The six-dimensional equation of motion for $F_\2$, namely 
$d(e^{\ft12\phi} {\bar *F_\2}) - e^\phi {\bar *\bar H_\3}\wedge F_\2=0$, 
reproduces the
four-dimensional Yang-Mills equations in (\ref{bHFieqns}).
The six-dimensional
Einstein equations should reproduce the four-dimensional 
equations found above, together with the four-dimensional Einstein 
equations.  They can all be derived from the four-dimensional Lagrangian
\be
{\cal L}_4 = R\, {*\oneone} -\ft12 {*d\phi}\wedge d\phi
   -\ft12 e^{2\phi}\, {*H_\3}\wedge H_\3 
    -\ft14 e^\phi\, {* F^i}\wedge F^i\,.
\ee
Note that if we send $A^i\rightarrow \sqrt2\, A^i$,  
$g\rightarrow \sqrt 2\, g$ and $\phi\rightarrow-\phi$, 
the $S^2$ Pauli reduction we have constructed here then corresponds to the
one in the conventions of \cite{gibpopss}.

\section{Conclusions}

  In this paper we have employed a relation between consistent
DeWitt (group manifold) reductions and Pauli (coset) reductions that was 
established in \cite{paulired}, applying it to two instances of $S^2$ 
reductions in supergravity theories.  In the first example, we addressed
the question of whether there exists a consistent $S^2$ Pauli reduction of 
five-dimensional minimal ungauged supergravity.  One might expect, in
view of the close parallels between the five-dimensional theory and
eleven-dimensional supergravity, that there could exist such a consistent
reduction, paralleling the known consistent $S^4$ reduction of the
eleven-dimensional theory.  Our starting point was the observation that
the minimal five-dimensional supergravity can be obtained as a 
consistent truncation of the five-dimensional supergravity
that one obtains from an $S^1$ Kaluza-Klein reduction of 
a minimal six-dimensional supergravity.  By performing a (necessarily
consistent) DeWitt reduction of the six-dimensional theory on the 
$S^3 =SU(2)$ group manifold, and then reducing this on the $U(1)$ 
Hopf fibres of the $S^3$, one obtains a (necessarily consistent) Pauli
$S^2$ reduction of the untruncated five-dimensional theory.  This does
not yet establish the consistency of the Pauli $S^2$ reduction of the
minimal five-dimensional supergravity, however, since for this to work
the truncation that is still needed in five dimensions would have to be 
compatible with the Pauli/DeWitt relation.  We showed that in fact one
cannot consistently perform the truncation of fields in five dimensions,
thus leading to the conclusion that a consistent Pauli $S^2$ reduction of
five-dimensional minimal supergravity is not possible.

   The second example we studied in this paper was concerned with the
known consistent Pauli $S^2$ reduction of the six-dimensional Salam-Sezgin
supergravity.  This reduction was derived by direct means \cite{gibpopss},
and thus although technically much simpler than other consistent Pauli
reductions such as the $S^4$ or $S^7$ reductions of eleven-dimensional
supergravity, the underlying reason for why it should exist remained
rather obscure.  Our aim in this paper was to try to gain an  
understanding of the consistency of the reduction by deriving it from a
manifestly-consistent DeWitt $SU(2)$ group manifold reduction from seven
dimensions.  A possible candidate for such an explanation was already
in existence, since it had been shown in \cite{cvgiposs} that the
Salam-Sezgin theory could indeed be derived via a consistent $S^1$
reduction, and truncation, of a seven-dimensional $SO(2,2)$-gauged
supergravity.  However, this reduction route was not suitable for
our present purposes, since the Kaluza-Klein vector in the reduction
from seven to six dimensions was in fact set to zero, and so it could not 
supply the needed ``twist'' that would promote the subsequent $S^2$ 
reduction into an $S^3$ reduction from seven dimensions rather than merely
$S^1\times S^2$. 

   We then showed that there exists a completely
different way of deriving the Salam-Sezgin theory\footnote{Or, at least,
its bosonic sector.  In our present considerations we have only been 
concerned with finding an explanation for the consistency of the Pauli
$S^2$ reduction of the bosonic sector of the Salam-Sezgin theory.  The
reduction of the fermionic sector will be addressed in \cite{azizi}.} as
a Kaluza-Klein $S^1$ reduction from the $SO(2,2)$-gauged seven-dimensional
supergravity, and in this construction the Kaluza-Klein vector is 
non-vanishing.  The lift of the $S^2$ reduction to seven dimensions indeed
now gives rise to an $S^3$ reduction, and so the possibility of
relating the Pauli and DeWitt reductions using the relations established
in \cite{paulired} arises.  The only remaining question is whether
the necessary truncation of fields in six dimensions to obtain the
pure Salam-Sezgin theory is compatible with the reduction of the DeWitt
reduction on its Hopf fibres.  In this case, unlike in the first example
we studied in this paper, we found that the truncation is compatible, and
so this provides an understanding of the consistency of the $S^2$ Pauli
reduction of the Salam-Sezgin theory.

\section*{Acknowledgments}

We are grateful to Hong L\"u and Ergin Sezgin for useful discussions.  
This work was supported in part by DOE grant DE-FG02-13ER42020.


\begin{thebibliography}{99}

\bibitem{strau} N. Straumann, {\it On Pauli's invention of non-Abelian
Kaluza-Klein theory in 1953}, gr-qc/0012054.

\bibitem{oraf} L. O'Raifeartaigh and N. Straumann, {\it Early history of
gauge theories and Kaluza-Klein theories, with a glance at recent
developments}, hep-ph/9810524.

\bibitem{dewnic} B. de Wit and H. Nicolai, {\it The consistency of the 
$S^7$ truncation in $D = 11$ supergravity},
Nucl. Phys. {\bf B281}, 211 (1987).

\bibitem{paulired} M. Cveti\v c, G.W. Gibbons, H. L\"u and C.N. Pope,
{\it Consistent group and coset reductions of the bosonic string},
Class.\ Quant.\ Grav.\  {\bf 20}, 5161 (2003), hep-th/0306043.

\bibitem{vann1} H. Nastase, D. Vaman and P. van Nieuwenhuizen,
{\it Consistent nonlinear KK reduction of 11-d supergravity on 
AdS$_7 \times S^4$ and selfduality in odd dimensions}, 
Phys.\ Lett.\ B {\bf 469}, 96 (1999), hep-th/9905075.

\bibitem{vann2} H. Nastase, D. Vaman and P. van Nieuwenhuizen,
{\it Consistency of the AdS$_7\times S^4$ reduction and the origin of  
self-duality in odd dimensions},
Nucl. Phys.  {\bf B581}, 179 (2000), hep-th/9911238.

\bibitem{cvluposatr} M. Cveti\v c, H. L\"u, C.N. Pope, A. Sadrzadeh 
and T.A. Tran,
{\it $S^3$ and $S^4$ reductions of type IIA supergravity},
Nucl.\ Phys.\ B {\bf 590}, 233 (2000), hep-th/0005137.

\bibitem{baghohsam} A. Baguet, O. Hohm and H. Samtleben,
{\it Consistent type IIB reductions to maximal 5D supergravity}, 
Phys.\ Rev.\ D {\bf 92}, no. 6, 065004 (2015), arXiv:1506.01385 [hep-th].

\bibitem{Gred} A. Baguet, C.N. Pope and H. Samtleben,
{\it Consistent Pauli reduction on group manifolds},
Phys.\ Lett.\ B {\bf 752}, 278 (2016), arXiv:1510.08926 [hep-th].

\bibitem{DeWitt} B.S. DeWitt, in {\it Relativity, groups and topology},
Les Houches 1963 (Gordon and Breach, 1964).

\bibitem{salsez} A. Salam and E. Sezgin,
{\it Chiral compactification on (Minkowski)$\times S^2$ of $N=2$ 
Einstein-Maxwell supergravity in six-dimensions},
Phys. Lett. {\bf B147}, 47 (1984).

\bibitem{gibpopss} G.W. Gibbons and C.N. Pope,
{\it Consistent $S^2$ Pauli reduction of six-dimensional chiral gauged 
Einstein-Maxwell supergravity},
Nucl.\ Phys.\ B {\bf 697}, 225 (2004), hep-th/0307052.

\bibitem{cvgiposs} M. Cveti\v c, G.W. Gibbons and C.N. Pope,
{\it A string and M-theory origin for the Salam-Sezgin model},
Nucl.\ Phys.\ B {\bf 677}, 164 (2004), hep-th/0308026.

\bibitem{lupose} H. L\"u, C.N. Pope and E. Sezgin,
{\it $SU(2)$ reduction of six-dimensional $(1,0)$ supergravity}, 
Nucl.\ Phys.\ B {\bf 668}, 237 (2003), hep-th/0212323.

\bibitem{azizi} A. Azizi, to appear.

\end{thebibliography}
\end{document}